\documentclass[preprint]{emulateapj}
\usepackage{hyperref}
\usepackage[usenames,dvipsnames]{color}
\hypersetup{colorlinks,citecolor=Blue,linkcolor=Red,urlcolor=Blue}
\usepackage{amsmath}

\begin{document}

\title{Magnetic Origins of the Stellar Mass-Obliquity Correlation in Planetary Systems}

\shorttitle{Magnetic Realignment} 
\shortauthors{Spalding \& Batygin} 
\author{Christopher Spalding*$^{1}$ and Konstantin Batygin$^{1}$} 

\affil{$^1$Division of Geological and Planetary Sciences, California Institute 
of Technology, Pasadena, CA 91125} 
\email{*cspaldin@caltech.edu}
\begin{abstract}
Detailed observational characterization of transiting exoplanet systems has revealed that the spin-axes of massive ($M\gtrsim1.2\,\textrm{M}_\odot$) stars often exhibit substantial misalignments with respect to the orbits of the planets they host. Conversely, lower-mass stars tend to only have limited obliquities. A similar trend has recently emerged within the observational dataset of young stars' magnetic field strengths: massive T-Tauri stars tend to have dipole fields that are $\sim$10 times weaker than their less-massive counterparts. Here we show that the associated dependence of magnetic star-disk torques upon stellar mass naturally explains the observed spin-orbit misalignment trend, provided that misalignments are obtained within the disk-hosting phase. Magnetic torques act to realign the stellar spin-axes of lower-mass stars with the disk plane on a timescale significantly shorter than the typical disk lifetime, whereas the same effect operates on a much longer timescale for massive stars. Cumulatively, our results point to a primordial excitation of extrasolar spin-orbit misalignment, signalling consistency with disk-driven migration as the dominant transport mechanism for short-period planets. Furthermore, we predict that spin-orbit misalignments in systems where close-in planets show signatures of dynamical, post-nebular emplacement will not follow the observed correlation with stellar mass.
\end{abstract}
\section{Introduction}
\label{sec:intro} 


%

Of all ideas in planetary science, few have stood the test of time better than the ``Nebular Hypothesis", originally proposed in the 18th century by Kant and Laplace \citep{Kant1755,Laplace1796}. The impetus for such a model was the near-aligned configuration of the planetary orbits with each other and with the Sun's spin axis; the net angular momentum of the the planets differs in direction from the Sun's spin by about $7\,\deg$\footnote{Of course, $7\,\deg$ is very different from zero and so is still in need of an explanation.} \citep{Beck2005}. Centuries of refinement (with help from astronomical observations) have resulted in the modern picture whereby a dense molecular cloud core collapses under its own self gravity \citep{Shu1987,Mckee2007} to form a star encircled by a disk of gas and dust. 

The earliest descriptions of molecular core collapse were naturally the most simplistic, supposing that the star and its disk both inherit similar angular momentum directions. More recent work \citep{Goodman1993,Caselli2002,Bate2010} has added a layer of complexity to the story by noting that turbulence in collapsing cores implies that the last bits of material falling onto the disk do not necessarily share the same angular momentum direction as the star. Despite this apparent tendency towards slight misalignment, the mutual gravitational torque between star and disk is likely strong enough during the earliest stages to stave off any significant star-disk misalignment arising from core turbulence \citep{Spalding2014b}.

Long before the first exoplanetary detections \citep{Mayor1995,Marcy1996}, the theory behind protoplanetary disks was already fairly mature (e.g., \citealt{Goldreich1979,Goldreich1980,Lin1986,Ward1986}). An early prediction from this field was that angular momentum exchange between disks and their embedded planets should give rise to planetary migration towards shorter-period orbits. More recent work has subjected this idea to extensive numerical \citep{Nelson2000,Rice2008} and analytic \citep{Tanaka2002} analyses, with the general picture of migration holding up as an expected outcome (for a recent review, see \citealt{Kley2012}). Sure enough, the earliest days of exoplanet-hunting revealed a considerable population of hot Jupiters, planets with about the mass of Jupiter but with orbital periods of a few days. Conventional planet formation theory \citep{Pollack1996} suggests that these planets must form at several AU, beyond the snow line, and must have migrated towards their present-day orbits. Accordingly, disk-driven migration seemed an attractive mechanism by which giant planets may be delivered to close-in orbits.

\begin{figure}[h!]
\centering
\includegraphics[trim=10.4cm 8cm 9cm 8cm, clip=true,width=1.1\columnwidth]{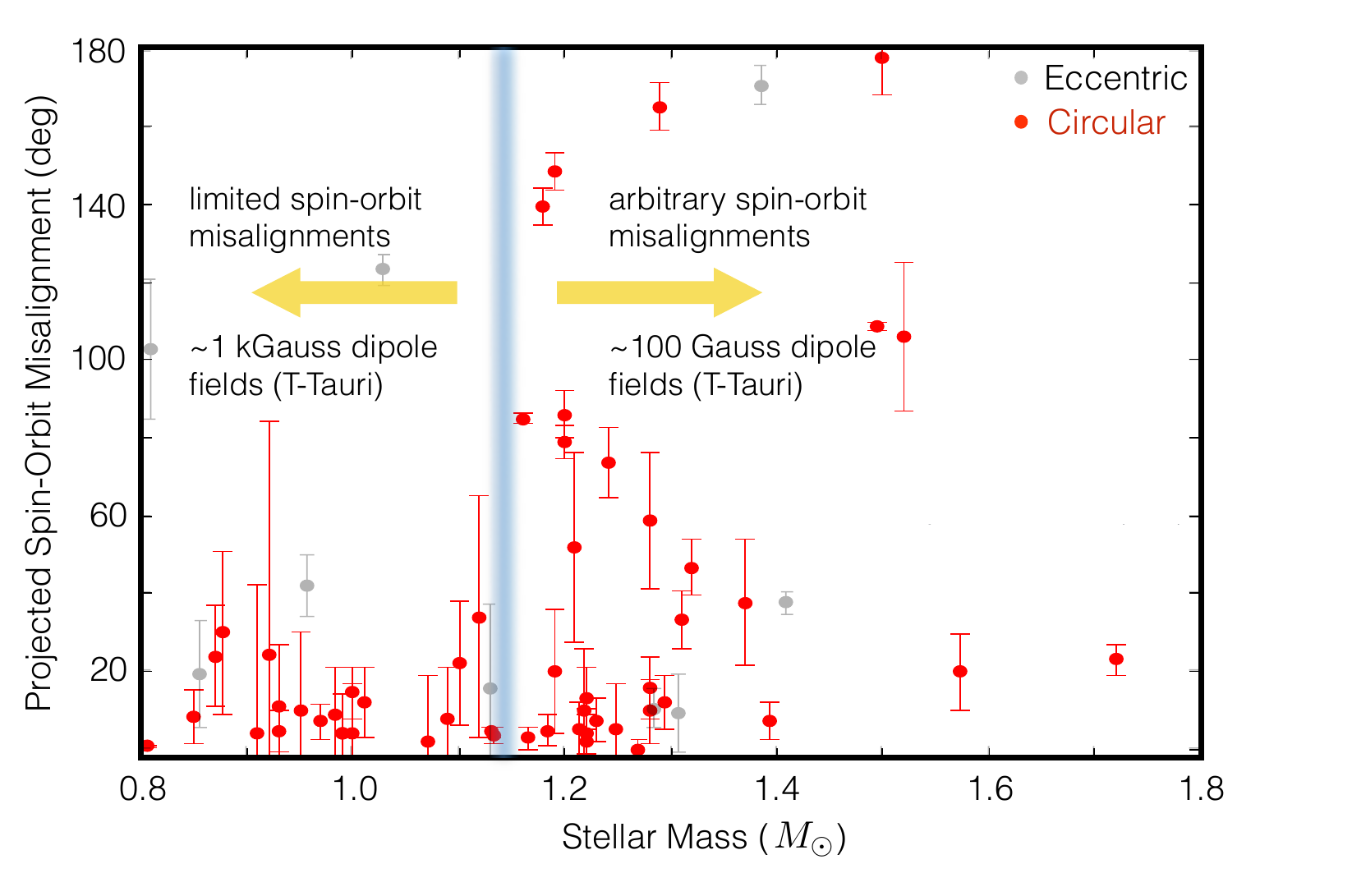}
\caption{The observed projected angle between the stellar spin axis and orbital plane of circular planetary orbits ($e\leq0.1$; red, solid points) and eccentric orbits ($e>0.1$; black, faint points) for stars of given masses. There exists a clear distinction between low-mass stars $(M\lesssim 1.2\,M_{\odot})$, which display moderate-to small misalignments (especially among circular systems), and the more massive stars $(M\gtrsim 1.2\,M_{\odot})$, which exhibit misalignments ranging all the way from retrograde-aligned to prograde-aligned. Measurements of magnetic field strengths \citep{Gregory2012} have revealed that, among T-Tauri stars, lower-mass stars (similarly, corresponding to $M\lesssim 1.2-1.4\,M_{\odot}$) possess a much stronger surface dipole field than do their higher-mass counterparts. Specifically, low-mass stars possess fields of $\sim$1\,kGauss in contrast to more modest $\sim$0.1\,kGauss for higher-mass T-Tauri stars. The misalignments data were obtained from \texttt{exoplanets.org} and follow the discussion of Albrecht et al. (2012).}
\label{Lambda}
\end{figure}

Until recently, it was impossible to tell whether these systems, supposed to migrate through a planar disk, were aligned or misaligned with their stars. This property, referred to variably as obliquity or spin-orbit misalignment, has now fallen within the observational capabilities of exoplanetary astronomy by way of the Rossiter-McLaughin effect \citep{Rossiter1924,McLaughlin1924,Winn2005}. Currently, measurements of misalignments are most common for the orbits of hot Jupiters, wherein the findings have revealed that a substantial fraction of such planets follow orbits with significant obliquities \citep{Winn2010,Albrecht2012}. Indeed, misalignments range all the way from prograde aligned to retrograde anti-aligned. However, the degree of misalignment exhibits a clear dependence on stellar mass, with the most extreme, retrograde (circular) orbits only appearing around stars with mass greater than $M\gtrsim1.2\,\textrm{M}_{\odot}$ (Figure~\ref{Lambda}).

In light of the existence of significant obliquities, many authors (e.g., \citealt{Albrecht2012,Lai2012,Dawson2014,Storch2014,Petrovich2015}) have sought alternatives to disk-driven migration as a production mechanism for inclined hot Jupiters. Specifically, in contrast to the `smooth' picture of disk-driven migration, there exists a separate class of migration mechanisms which occur after the dissipation of the protoplanetary disk. Post-disk migration must invoke dynamical interactions to excite near-unity eccentricities, bringing the periastron close enough to the central star such that tidal forces may take over and circularize the orbit \citep{Ford2006}. The necessary perturbations are hypothesized to arise from processes such as planet-planet scattering \citep{Ford2008,Nagasawa2008,Beauge2012}, Kozai resonance with a perturbing companion \citep{Wu2001,Fabrycky2007,Naoz2011} or secular chaotic excursions \citep{Lithwick2012}.  

 Despite the natural tendency of dynamical interactions to excite inclinations, their occurrence rate appears to be insufficient to explain the current data \citep{Dawson2015}. Conversely, disk-driven migration is expected to be almost ubiquitous but has long been assumed to give rise to low-obliquity systems because of the tendency for an isolated collapsing core to form an aligned star-disk system. However, stars do not form in isolation. On the contrary, binary fraction probably lies somewhere between half and unity during the early, disk-hosting stage of stellar evolution \citep{Ghez1993,Kraus2011,Marks2012,Duchene2013}. The potential for neighbouring stars to influence disk orientation has actually been recognized for some time (e.g. \citealt{Larwood1996}), but it was first suggested as an explanation for spin-orbit misalignments by \citet{Batygin2012}. In this scenario, the gravitational torque exerted by a distant companion star causes the disk to precess about the plane of the binary system. Observations suggest that binary planes do not correlate with disk orientation \citep{Stapelfeldt1998,Koresko1998,Jensen2014} and so the disk-precession tends to torque the star and disk out of alignment. 

We recently advanced the so-called ``disk-torquing" framework by taking into account the gravitational torques communicated between the star and its disk \citep{Spalding2014}. Specifically, the star tends to precess about the disk's plane at a rate that decreases with time owing both to stellar contraction and disk mass-loss. In the earliest stages, the star is well enough coupled to the disk to adiabatically trail the disk as it precesses. However, once the precession rate of the star about the disk is roughly commensurate with that of the disk about the binary plane, a secular resonance is encountered which is capable of exciting star-disk misalignments occupying the entire observed range. Such a picture has been independently corroborated in other work and appears to be a robust result across various levels of approximation \citep{Batygin2013,Lai2014,Xiang-Gruess2014}.

From the above discussion, it appears that spin-orbit misalignments may arise naturally out of either class of proposed migrationary scenario. What is less obvious is how to reconcile the migration pathway with the mass-dependence of the obliquity distribution (Figure~\ref{Lambda}). Previous attempts to explain the trend \citep{Winn2010,Lai2012} have largely focussed on the violent migration pathway. Specifically, it has been proposed that dynamical encounters misalign orbits, after which, strong tidal coupling between low-mass stars and their planets (owing to an extended convective region) re-aligns the orbits. These hypotheses rest upon a number of assumptions, one of which being that the tidal damping rate of obliquity greatly exceeds that of orbital semi-major axis \citep{Lai2012}. Additionally, the presence of misaligned multi-transiting systems \citep{Huber2013} has yet to be given a viable explanation within the violent framework. 

The disk-torquing mechanism is expected to misalign all planets in the system by a similar amount, at least in the inner disk. However, to date, the mass-misalignment trend has not been given an adequate explanation. Here, we provide such an explanation by utilizing an additional piece of evidence in the form of the magnetic field strengths of young, disk-hosting stars. Specifically, recent observations \citep{Gregory2012} have revealed that low-mass T-Tauri stars possess dipole field strengths of order $\sim1$\,kG, an order of magnitude greater than their higher-mass counterparts ($\sim0.1$\,kG). This trend appears to continue into the more massive Herbig AeBe class of star-disk systems ($M\sim1.5-8\,\textrm{M}_{\odot}$) \citep{Alecian2013}, suggesting the dipole strength to be a robust mass-dependent feature, intrinsically connected to PMS evolution. Crucially, the transition from weaker to stronger fields occurs at a similar mass to that defining the mass-misalignment trend ($\sim1.2\,\textrm{M}_{\odot}$). In this work, we suggest that the two trends are causally linked: \textit{stronger magnetic fields of low-mass stars erase primordial misalignments that their higher-mass, weaker-field counterparts retain}. 

 In order to test this idea, we must first quantify the torques communicated between a tilted dipole and a disk. We are not the first to suggest that magnetic torques might influence misalignments \citep{Lai2011} but our goal is to couple the star-disk magnetic torques to the full disk-torquing picture, i.e., a freely precessing star-disk system. Consequently, we develop a more complete picture of the various torques involved using semi-analytic methods, whilst taking advantage of the conclusions of various numerical studies (mostly on aligned disks, e.g. \citealt{Ghosh1978,Armitage1996,Uzdensky2002}). Highly sophisticated numerical models exist for the study of star-disk interactions (e.g. \citealt{Romanova2012}), however, detailed simulations must be integrated in full throughout each stellar rotation period. Even at the current state-of-the-art, full 3D calculations can only be carried out within a reasonable length of computer time for $\sim1000$ orbital timescales (depending upon the precision). In contrast, we seek to model the global evolution of the star-disk system, including time-dependent disk mass and stellar contraction, over multi-Myr timescales. Considering also that current observations only constrain the topologies of T-Tauri field strengths to within an order of magnitude or so, we shall construct our model such that the level of detail is commensurate with that of the observations.

In this paper, we begin with a description for the time evolution of the star and disk. Next we provide a re-hashing of the purely gravitational torques as derived in \citet{Spalding2014}, derive the various magnetic torques communicated between the disk and star, and subsequently present the results obtained through numerical integration of the star-disk system, complete with gravitational and magnetic torques. We conclude by discussing the implications of the result for the acquisition of spin-orbit misalignments within the disk-torquing framework.

\section{Model}

Our goal is to describe the spin-axis dynamics of a star, possessing a dipolar field that is encircled by a protoplanetary disk. We suppose the star-disk system to be orbited by a companion star. In what follows, we derive the analytical forms of the various torques inherent to the system. Throughout the entire calculation, we assume a hierarchical configuration. Specifically, we assume that the central star and binary companion do not influence each other directly. Rather, the companion torques the disk (gravitationally), which in turn interacts with the central star (gravitationally and magnetically). In order to improve clarity, we adopt the following convention for variables with identical symbols: quantities referring to the disk are primed, those referring to the central star are marked with a tilde, and those referring to the companion star are given an over-bar (e.g. semi-major axes will be defined as $a'$, $\tilde{a}$ and $\bar{a}$ respectively).

The binary companion is prescribed a mass $\bar{M}$ and an orbit with semi-major axis $\bar{a}$ which is much greater than the outer disk edge $a'_{\textrm{out}}\approx30\,$AU. The resulting gravitational torques acting upon the disk cause it to precess at a frequency that depends upon the mass of the companion, its semi-major axis and its inclination relative to the disk-plane \citep{Spalding2014}. In the absence of star-disk interactions, the precession of the disk would simply cause it to tilt out of alignment with its host star. However, the central star can interact with its disk by way of several physical processes. Specifically, young stars rotate rapidly enough to possess a considerable centrifugal bulge at their equators. This bulge allows for gravitational coupling between the disk and star, the dynamics of which having been derived in detail elsewhere \citep{Batygin2013,Spalding2014}. In addition, observations of young stars have revealed the presence of significant magnetic fields \citep{Johns-Krull2007,Gregory2012} which facilitate angular momentum transfer between star and disk in addition to the gravitational influences.

 Our ultimate goal is to develop a theoretical framework by which we may test the hypothesis that differences in stellar magnetic field between high and low-mass stars is the dominant driver of the observed mass-misalignment trend in the current Rossiter-McLaughlin dataset \citep{Winn2010,Albrecht2012}. We follow a semi-analytic, parameterized framework in deriving the relevant equations. 
 
  As the system ages, the star contracts and the disk loses mass. Accordingly, before calculating the star-disk torques, we must first provide formulations of the physical evolution of the disk and the central star. Within such a framework we then describe our calculations of magnetically-facilitated tilting of the stellar-spin axis and star-disk gravitational coupling. Torques arising from accretion are neglected in this work because their influence within a similar physical framework has been examined elsewhere \citep{Batygin2013} and found to be insignificant.  

\subsection{Physical Evolution of the Protoplanetary Disk and the Stellar Interior}

Typically quoted lifetimes of protoplanetary disks fall in the range $\sim 1-10$\,Myr, with some recent evidence in support of longer-lived systems \citep{Haisch2001,Williams2011,Bell2013}. We parameterize disk mass evolution as \citep{Laughlin2004}: 

\begin{equation}
M_{\rm{disk}} = \frac{M_{\rm{disk}}^0}{1 + t/\tau_{\rm{disk}}}. 
\end{equation}

 \noindent The time derivative of $M_{\rm{disk}}$ approximately represents the accretionary flow. Following \citet{Spalding2014}, we note that observations \citep{Hartmann2008,Herczeg2008,Hillenbrand2008} are best matched by adopting an initial disk mass, $M_{\rm{disk}}^0 = 5 \times 10^{-2}\,\textrm{M}_{\odot}$ and dissipation timescale $\tau_{\rm{disk}} = 5 \times 10^{-1}$\,Myr. 

For simplicity, we approximate the central star with a polytrope of index $\xi=3/2$ (appropriate for a fully convective object; \citealt{Chandrasekar1939}). A polytropic body of this index possesses a specific moment of inertia $I = 0.21$ and a Love number (twice the apsidal motion constant) of $k_2 = 0.14$. In contracting along their respective Hyashi Tracks, T-Tauri stars derive most of their luminosity from the release of gravitational potential energy. We describe the radiative loss of such binding energy as \citep{Hansen1994}:

\begin{equation}\label{Contract}
- 4 \pi R_\star^2 \sigma T_{\rm{eff}}^4 =  \left( \frac{3}{5-\xi} \right) {G M_\star^2 \over 2 R_\star^2} {d R_\star \over dt},
\end{equation}

with a solution,

\begin{equation}\label{Rstar}
R_{\star} = (R_{\star}^0) \left[1+ \left( \frac{5-\xi}{3} \right) \frac{24 \pi \sigma T_{\rm{eff}}^4}{G M_{\star} (R_{\star}^0)^3} t \right]^{-1/3}.
\end{equation}

For definiteness, we match the numerical evolutionary track of an $M_{\star} = 1\,\rm{M}_{\odot}$ star \citep{Siess2000} by assuming an initial radius of $R_{\star}^0 \simeq 4 R_{\odot}$ and an effective temperature of $T_{\rm{eff}} = 4100$\,K.

\subsection{Disk-Binary Gravitational Interactions}

The binary companion interacts with the disk through gravitational torques alone. To capture the long-term behaviour of the angular momentum exchange, we utilize the secular approximation \citep{Murray2000,Morby2002} whereby the torques exerted by the companion on the disk are equivalent to those communicated by a massive wire sharing the companion's orbital elements. As a consequence of self-gravity and hydrodynamic pressure forces, the disk retains coherence under the influence of such torques, acting as a rigid body \citep{Larwood1996,Batygin2011,Xiang-Gruess2014}. Additionally, we neglect any dynamical influence of disk-warping.

We adopt a Hamiltonian framework in describing the dynamics. The gravitational torques acting between a star, disk and companion have already been comprehensively derived and discussed elsewhere \citep{Spalding2014}. Accordingly, here we provide only a brief outline of the corresponding computation. 

To begin, we introduce the scaled Poincar$\acute{\rm{e}}$ action-angle coordinates, defined in terms of the disk-star mutual inclination $\beta'$ and the disk's argument of ascending node $\Omega'$,

\begin{equation}
\label{scalepoinc}
Z'=1-\cos(\beta') \ \ \ \ \ \ \ z' = -\Omega'.
\end{equation} 

The appropriate Hamiltonian describing the companion's influence upon the disk is then

\begin{align}
\mathcal{U}&= \frac{3 n'_{\rm{out}}}{8} \frac{\bar{M}}{M_{\star}} \left(\frac{a'_{\rm{out}}}{\bar{a}} \right)^3 \left[ Z'-\frac{Z'^2}{2} \right],
\label{HUpsilon}
\end{align}

\noindent where $n_{\textrm{out}}$ is the mean motion of the gas at the outer edge of the disk. Crucially, $\mathcal{U}$ contains no explicit dependence on $z'$. Therefore, \textit{the disk-binary inclination is a constant of motion}. Appropriately, in our analysis, we carry out our simulations within a reference frame co-precessing with the disk, at a rate $\nu = dz'/dt$, given by 

\begin{equation}
\nu = \frac{\partial \mathcal{U}}{\partial Z'} = \frac{3 n'_{\rm{out}}}{8} \frac{\bar{M}}{M_{\star}} \left(\frac{a'_{\rm{out}}}{\bar{a}} \right)^3 \bigg[ \,1- Z' \bigg] .
\end{equation}

\noindent Boosting into such a frame is equivalent to subtracting $\nu t$ from the argument of ascending node of the disk.
 
Owing to the arbitrary nature of choosing the companion's orbital parameters, we prescribe $\nu$ and $Z'$ in our problem independently. It is worth noting that the picture whereby a single companion remains on a circular orbit throughout the entire disk lifetime is highly idealized. In reality there may exist multiple companions and/or companions might be gained and lost throughout the pre-main sequence. Of course, these complications will lead not only to a time-dependent $\nu$, but also to secular variations in $Z'$. This point is not crucial to the problem at hand, so we leave analysis of such dynamic processes to future work and maintain a constant $\nu$ and $Z'$ throughout.

\subsection{Disk-Star Interactions: Gravity}

Having prescribed the secular evolution of the disk owing to the companion (constant, rigid-body precession), we now describe the processes by which torques are communicated between the star and the disk. Observations of T-Tauri stars have revealed that they spin with periods ranging between $\sim3-10$\,days \citep{Herbst2007,Littlefair2010,Bouvier2013}. Such high spin rates ($\sim$within a factor of ten of break-up rotation) lead to the development of a significant centrifugal bulge on the stellar equator. When the star and disk are misaligned, this bulge results in gravitational torques that force a precession of the stellar spin pole about the disk plane (analogous to a top spinning on a planar table). 

 The dynamics of a spheroidal star and the gravitational influence arising from its rotationally-derived equatorial bulge may be approximated to high precision by considering the inertially-equivalent picture of a wire with mass $\tilde{m}$ in circular orbit with semi-major axis $\tilde{a}$ around a point mass. Respectively, the appropriate mass and semi-major axis are given by \citep{Batygin2013}
 
\begin{align}
\tilde{m} &= \left[ 
\frac{3 M_{\star}^2 \omega^2 R_{\star}^3 I^4}{4 G k_2 } \right]^{1/3},\nonumber\\
\tilde{a} &= \left[
\frac{16 \omega^2 k_2^2 R_{\star}^6}{9 I^2 G M_{\star}} \right]^{1/3}.
\label{astaryo}
\end{align} 

\noindent  With this prescription, the standard perturbation techniques of celestial mechanics can be applied to the spheroidal star \citep{Murray2000,Morby2002}.

By working in a frame co-precessing with the disk, we introduce a time-dependence to the Hamiltonian resulting in an apparent linear increase in the disk's argument of ascending node of $\nu\,t$. The Hamiltonian can be made autonomous by employing a canonical transformation arising from the following generating function of the second kind \citep{Goldstein1950}:

\begin{equation}
\mathbb{G}_2 = (\tilde{z}-\nu t)\,\Phi,
\end{equation}

\noindent where $\phi = (\tilde{z}-\nu t)$ is the new angle and the new momentum is related to the old one through:

\begin{equation}
\tilde{Z} = \frac{\partial \mathbb{G}_2}{\partial \tilde{z}} = \Phi.
\end{equation}

In addition to removing explicit time dependence, we scale the Hamiltonian by $\nu$. Following the transformations described above, the Hamiltonian takes on the following form:

\begin{align}
\mathcal{H} &= - \Phi + \frac{\tilde{\delta}}{12} \bigg[ 3 \big(\Phi -2 \big) \Phi\nonumber\\
& - 3 \big(2+3(\Phi -2)\big) \Phi \cos^2(\beta') \nonumber \\ 
&+ 6 \sin(2 \beta') \big(\Phi -1\big) \sqrt{(2-\Phi) \Phi} \cos(\phi)\nonumber\\
 &+ 3 \sin^2(\beta')\big(\Phi -2\big) \Phi \cos(2 \phi) \bigg].
\label{HammyK}
\end{align}

\noindent where, 

\begin{eqnarray}\label{delta}
\tilde{\delta}\equiv\frac{3}{8}\left(\frac{{n'}_{\rm{in}}^2}{\omega \nu}  \frac{M_{\rm{disk}}}{M_{\star}} \frac{a'_{\rm{in}}}{a'_{\rm{out}}} \right),
\end{eqnarray}

\noindent and $n'_{\textrm{in}}$ is the mean motion at the disk's inner edge.

 The purely gravitational dynamics described by the above Hamiltonian (\ref{HammyK}), together with the physical evolution of the star and disk, give rise to the excitement of mutual inclination between the central star and its disk \citep{Spalding2014}. However, additional physical mechanisms must exist in order to explain the mass-misalignment trend. The main hypothesis of our paper is that the dominant driver of such a trend is the mass-dependence of T-Tauri dipole field strengths. Accordingly, next we present our derivations of the magnetic torques and the differential equations used in determining their influence on stellar spin dynamics. 

\subsection{Magnetic Torques}

In order to prescribe the magnetic disk-star interactions, we consider a T-Tauri star possessing a purely dipole magnetic field, whose north pole is aligned with the stellar spin axis. A pure dipole is modeled because the octupole component in real systems falls off much faster with distance ($\propto 1/r^5$) than the dipole ($\propto 1/r^3$). Accordingly, at the position of the inner edge of the disk ($\sim10\,R_\star$), the octupole components have been attenuated to a factor of $\sim 10^{-5}$ relative to the stellar surface field ($r=R_\star$) as opposed to the $\sim 10^{-3}$ attenuation suffered by the dipole component. Observations constrain the surface octupole field to differ from the dipole field by little more than a factor of 10 (higher or lower) and so the octupole component is almost always negligible when considering disk-star torques. The stronger dipoles of low-mass stars directly lead to a greater magnetic interaction with their disks than for high-mass stars.

 In the region of interest (i.e. in the domain of the disk), the stellar field is current-free and can be expressed as the gradient of a scalar potential:
 
\begin{equation}
\mathbf{B}_{\rm{dip}}=-\boldsymbol{\nabla} V.
\end{equation}

To retain generality, we take the field to be tilted at an angle $\beta'$ with respect to a spherical coordinate system ($r\,,\,\theta\,,\,\phi$) into a direction specified by an azimuthal angle $\psi$:

\begin{align}
V&=B_{\star}R_\star \bigg(\frac{R_\star}{r}\bigg)^2 \bigg[ P_0^1(\cos(\theta)) \cos(\beta')\nonumber \\
&-\sin(\beta')\bigg(\sin(\psi)\sin(\phi) \nonumber \\
&+  \cos(\psi)\text{cos}(\phi)\bigg)P_1^1(\text{cos}(\theta)) \bigg],
\end{align}

\noindent where $B_{\star}$ is the equatorial stellar surface field and $P_l^m$ are associated Legendre polynomials. Within such a framework, the tilted dipole components are as follows,

  \begin{align}\label{dipole}
B_r&=2\,B_\star \bigg(\frac{R_\star}{r}\bigg)^3\big(\cos{\beta'}\cos(\theta)\nonumber\\
&+\cos(\phi-\psi)\sin(\beta')\sin(\theta)\big)\nonumber\\
B_\theta&=B_\star \bigg(\frac{R_\star}{r}\bigg)^3\big(\cos{\beta'}\sin(\theta)\nonumber\\
&-\cos(\phi-\psi)\sin(\beta')\cos(\theta)\big)\nonumber\\
B_\phi&=B_\star \bigg(\frac{R_\star}{r}\bigg)^3\big(\sin(\beta')\sin(\phi-\psi)\big).
\end{align}

\noindent  which describe the vector field
  
  \begin{equation}
  \mathbf{B}_{\textrm{dip}}=B_r \mathbf{\hat{e}}_r+B_\theta\mathbf{\hat{e}}_\theta+B_\phi \mathbf{\hat{e}}_\phi.
  \end{equation}
  

If we assume that the disk material is in Keplerian orbit about the star and for now suppose the star and disk are aligned, there exists a well-known expression for the corotation radius  $a'_{\rm{co}}|_{\beta'=0}=(\,G\,M_{\star}/\omega_{\star}^2)^{\frac{1}{3}}$, the radius at which relative angular velocity between the stellar magnetosphere ($\omega_{\star}$) and the disk material is zero. Now consider tilting the star with respect to the disk. The azimuthal motion of the stellar magnetosphere at the disk plane becomes reduced in such a way that the corotation radius is modulated as follows

\begin{equation}\label{coro}
a'_{\textrm{co}}=\bigg(\frac{G\,M_{\star}}{\omega^2_{\star}\,\cos^2(\beta')}\bigg)^{\frac{1}{3}}.
\end{equation} 

\noindent In other words, \textit{as the star tilts, the proportion of the disk which is effectively super-rotating relative to stellar rotation increases}. At radii greater or smaller than the corotation radius, Keplerian shear results in relative motion between the stellar magnetosphere and the fluid comprising the disk (Figure~\ref{Torques}).

 Owing to thermal ionization of alkali metals within the inner regions of the disk, motion of disk material relative to the stellar magnetosphere `drags' the field lines, inducing magnetic fields. The evolution of induced fields is governed by the induction equation,
 
\begin{equation}\label{Induct}
\frac{\partial \mathbf{B}}{\partial t}=\nabla \times \bigg(\eta \nabla \times \mathbf{B} + \mathbf{v}\times \mathbf{B} \bigg),
\end{equation}

\noindent where $\mathbf{B}$ is the total magnetic field,  $\eta$ is the magnetic diffusivity and

\begin{equation}
\mathbf{v}=v_K\,\mathbf{\hat{e}}_{\phi}-\boldsymbol{\omega}\times\mathbf{r}
\end{equation}

\noindent is the motion of the disk fluid \textit{in the frame rotating with the stellar magnetosphere}, i.e., we imagine the stellar magnetosphere to be held fixed, with the disk moving within it. The Keplerian velocity $v_K$ is azimuthally-directed whereas stellar rotation, arising from stellar angular velocity $\boldsymbol{\omega}$ has both azimuthal and vertical components. 
 
A full, time-dependent solution of equation~(\ref{Induct}) is computationally difficult and so we adopt an approximate, semi-analytic approach. Specifically, we simplify the picture by noting that the disk fluid velocity $\mathbf{v}$ possesses two separate components in the rotating frame. The first is an azimuthal component, arising from motion of disk material in Keplerian orbit relative to the azimuthal magnetospheric motion from stellar rotation. Second, there is a vertical component, where `vertical' refers to normal to the disk's plane. Vertical motion occurs owing to the tilt of the star with respect to its disk, which causes the stellar magnetosphere to be dragged vertically through the disk every stellar rotation period (best imagined for a star tilted by 90\,$\deg$; red diagram in Figure~\ref{Torques}). In a frame rotating with the star, this vertical dragging appears as a vertical motion of disk material (a vertical component to $\mathbf{v}$) and therefore constitutes an additional source of magnetic induction (see section~\ref{Paddle}). 

 For a small region on either side of the corotation radius, the relative azimuthal motion may be sufficiently small to allow a steady state to exist between magnetic field dragging and slippage \citep{Matt2004}. However, everywhere outside of this small region (of annular thickness $\sim 0.01\,a'_{\textrm{co}}$), the diffusive timescale is longer than the dragging timescale, leading to the unbounded inflation of magnetic field lines. Such inflation cannot physically continue indefinitely and so some mechanism must act to dissipate magnetic energy on orbital timescales. 
 
 Analytic and numerical models of the aligned star-disk configuration find that the most likely dissipative process is magnetic reconnection, whereby the magnetic field lines are stretched azimuthally until the induced field is of the same order as the vertical stellar field. At this point the magnetic field lines break and reform \citep{Livio1992,Uzdensky2002}. Such a process is intrinsically non-steady, however. Owing to the short timescale of reconnection ($\sim$ the orbital period) compared to disk evolution, we may average over each orbit and consider a steady magnetic torque to result between the star and disk.  
 
 We shall proceed by considering separately the fields generated from azimuthal relative motion and those originating from vertical (in the disk's frame) relative magnetospheric motion. The form of equation~(\ref{Induct}) ensures that the fields induced by vertical and azimuthal velocities back-react upon each other (through the $\mathbf{v}\times\mathbf{B}$ term) and so such separation of vertical and azimuthal induction is not strictly accurate. However, as we are seeking an approximate model for the magnetic star-disk interactions, we proceed with the picture whereby the two components act independently. 
 

\subsubsection{Azimuthal Induction}

Relative azimuthal disk motion induces an azimuthal field\footnote{Physically, the motion of the disk material relative to the background field generates a current within the disk. For purely-azimuthal fluid motion, the current is radial. This radial current in turn induces its own an azimuthal magnetic field (in the ideal MHD case), which leads the background field lines to appear stretched.} through flux-freezing \citep{Armitage1996,Agapitou2000}. This induced field, $B_{\phi,\textrm{i}}$ is represented as a fraction (also called a pitch angle), $\gamma$ of the component of the stellar dipole field perpendicular to the disk's surface. In the case of a thin disk in spherical coordinates, the `vertical' field is well approximated by the (negative) $\theta-$component of the stellar field, $B_{\theta}$, at the disk mid plane ($\theta=\pi/2$). Thus, $B_{\phi,\textrm{i}}=\gamma B_\theta$, where the subscript `i' refers to `induced'.

 As mentioned above, $\gamma$ is unable to instantaneously greatly exceed unity, but rather, magnetic reconnection reduces $\gamma$ to $\gamma\sim0$ each orbital period, allowing the field to be re-wound. We average over each reconnection timescale and consider the star-disk torque to act equivalently to a steady torque of azimuthal pitch angle $\gamma\sim1$. The force per unit volume is given by the Lorentz equation

\begin{equation}\label{Lorentz}
\mathbf{f}=\mathbf{J}\times\mathbf{B}.
\end{equation}

In the case of azimuthal induction, the current density $\mathbf{J}=(1/\mu_0)\nabla \times \mathbf{B}_{\textrm{ind}}$ may be considered to arise from the variation in magnetic field going from outside the disk vertically into the plane of the disk over some length scale $\delta$. In such a case, the current induced is radial and within the plane of the disk. Effectively, this statement is equivalent to setting the $\nabla$ operator equal to a vector of magnitude $1/\delta$ in the vertical direction. Assuming $\delta\ll h$ where $h$ is the disk scale height, the final torque per unit area on the disk is obtained approximately by multiplying equation~(\ref{Lorentz}) by $\delta$, effectively integrating the torque over the vertical dimension. 

Angular momentum transport among neighbouring annuli of the disk is facilitated by the propagation of bending waves \citep{Foucart2011}, disk viscosity \citep{Larwood1996} and through disk self-gravity \citep{Batygin2011}, all of which generally occur over a shorter timescale than does stellar tilting. As a result, the effective moment of inertia of the disk in response to magnetic torques is much greater than that of the star, as was the case above in the gravitational picture. Accordingly, in the calculations which follow, we consider only the back-reaction of the torques, in determining the dynamics of the star (which effectively introduces a minus sign), with the disk's dynamics being forced solely by the perturbing companion. 

From considering only the azimuthal field induction, arising from the penetration of stellar flux into the upper and lower surfaces of the disk, a torque per unit area of 

\begin{equation}\label{perArea}
\boldsymbol{\tau}=-\mathbf{r}\times\bigg[\gamma B_{\theta}\mathbf{\hat{e}}_r\times\mathbf{B}_{\textrm{dip}}\big|_{\theta=\pi/2}\bigg]
\end{equation}

\noindent is communicated between the disk and the star. It is appropriate to display these torques in terms of their Cartesian components in the disk frame (i.e. $\mathbf{\hat{e}}_z$ points along the disk angular momentum axis) before integrating over the entire disk. We find that the $\mathbf{\hat{x}}$, $\mathbf{\hat{y}}$ and $\mathbf{\hat{z}}$ torques arising from equation~(\ref{perArea}) within the region interior to corotation ($a<a_{\textrm{co}}$) are given by,

 \begin{align}\label{Interior}
 \tau_x &=\frac{2\,\pi\,\mathcal{T}}{3}\frac{( a_{\textrm{co}}^3-a_{\textrm{in}}^3)}{a_{\textrm{co}}^3a_{\textrm{in}}^3}\sin(\beta)\cos(\beta)\cos(\psi)\nonumber\\
 \tau_y &=\frac{2\,\pi\,\mathcal{T}}{3}\frac{( a_{\textrm{co}}^3-a_{\textrm{in}}^3)}{a_{\textrm{co}}^3a_{\textrm{in}}^3}\sin(\beta)\cos(\beta)\sin(\psi)\nonumber\\
 \tau_z &=\frac{4\,\pi\,\mathcal{T}}{3}\frac{( a_{\textrm{co}}^3-a_{\textrm{in}}^3)}{a_{\textrm{co}}^3a_{\textrm{in}}^3}\cos(\beta)^2,
 \end{align}
 
\noindent  where the variable $\mathcal{T}$, a measure of stellar magnetic moment, is defined for ease of writing as
 
 \begin{equation}
\mathcal{T}\equiv \frac{B_{\star}^2\,R_{\star}^6}{\mu_0}.
\end{equation}

 Analogous torques generated by the region exterior to corotation are similar in functional form but instead of integrating the torques between $a_{\textrm{in}}$ and $a_{\textrm{co}}$, we integrate outwards from $a_{\textrm{co}}$ and note that the outer edge of the disk is sufficiently large as to be approximately equivalent to integration out to $a'\rightarrow \infty$. Accordingly, the torques arising from outside of corotation are given by
 
  \begin{align}
 \tau_x &=-\frac{2\,\pi\,\mathcal{T}}{3}\frac{1}{a_{\textrm{co}}^3}\sin(\beta)\cos(\beta)\cos(\psi)\nonumber\\
 \tau_y &=-\frac{2\,\pi\,\mathcal{T}}{3}\frac{1}{a_{\textrm{co}}^3}\sin(\beta)\cos(\beta)\sin(\psi)\nonumber\\
 \tau_z &=-\frac{4\,\pi\,\mathcal{T}}{3}\frac{1}{a_{\textrm{co}}^3}\cos(\beta)^2.
 \end{align}

It is important to notice that the above torques are null for a star-disk inclination of $\beta=\pi/2$. The torque vanishes in such a scenario because in the picture considered thus far, at $B_\theta=0$ no flux penetrates the upper and lower surfaces of the disk. In other words, the null torque arises as an artifact of assuming a razor-thin disk. In reality, the disk possesses a finite scale height $h$. For small $\beta$, most of the stellar magnetospheric flux penetrates the upper and lower surfaces of the disk, validating the razor-thin model. However, as $\beta$ is increased, more stellar flux penetrates the inner edge of the disk. This flux penetration leads to additional azimuthal twisting of field lines within a small annular region at the disk truncation radius (Figure~\ref{Torques}).  

Whereas the induction considered previously generated a radial current, the induction owing to horizontal penetration of flux in the inner disk wall causes the field strength to vary along the radial direction. By the curl operator in equation~(\ref{Induct}), radial variation loosely translates to the generation of a vertical current. If we again suppose that the induced field is of a similar magnitude to the background stellar field, the current generated is that given by a change in magnetic field strength of $B_{r}(r=a_{\textrm{in}})$ (the radial stellar field evaluated at the truncation radius) over a radial length scale $\delta_r$. Once again, we integrate over this small region, assuming $\delta_r/a_{\textrm{in}}\ll1$, to find a torque per unit area (where the area is now the inner face of the disk) of 
\begin{equation}
\boldsymbol{\tau}_r=\frac{1}{\mu_0}a_{\textrm{in}}\mathbf{\hat{e}}_r\times\bigg[-B_r\mathbf{\hat{e}}_\theta \times\mathbf{B}_{\textrm{dip}}\big|_{r=a_{\textrm{in}}}\bigg].
\end{equation}

  Once again, converting to Cartesian components and integrating, we obtain the torques arising from radial flux penetration in the form,

 \begin{align}\label{Inner}
 \tau_x &=0\nonumber\\
 \tau_y &=0\nonumber\\
 \tau_z &=\frac{8\,\pi\,\alpha\mathcal{T}}{a_{\textrm{in}}^3}\sin(\beta)^2\nonumber\\
 \end{align}
 
 \noindent where $\alpha\equiv h_{\textrm{in}}/a_{\textrm{in}}$ is the aspect ratio evaluated at the inner edge of the disk. Its value is likely to be about $\sim0.1$ \citep{Armitage2011}, although it may indeed be slightly larger at the inner disk edge as a result of thermal expansion from strong stellar irradiation. For definiteness, we set $\alpha=0.1$ in our calculations.

\subsubsection{Vertical Induction}\label{Paddle}
Many previous works have considered the torques arising from azimuthal field dragging within accretion disks \citep{Ghosh1978,Livio1992,Armitage1996,Agapitou2000,Uzdensky2002,Matt2004,Lai2011}. However, a feature which has been omitted from previous works is the fact that if a stellar magnetosphere is rotating at an inclination relative to its disk, there exists a component of relative star-disk motion which forces the field lines to be dragged \textit{vertically} through the disk. This process is best imagined in the case of a star inclined by $\pi/2$ to a disk, i.e., spinning on its side. In such a case, consider the situation in a frame co-rotating with the star. In this frame, the disk is being forced to push through and break all stellar field lines each rotation period. 

\begin{figure*}[h!t]
\centering
\includegraphics[trim=0cm 2cm 0cm 0cm, clip=true,width=1\textwidth]{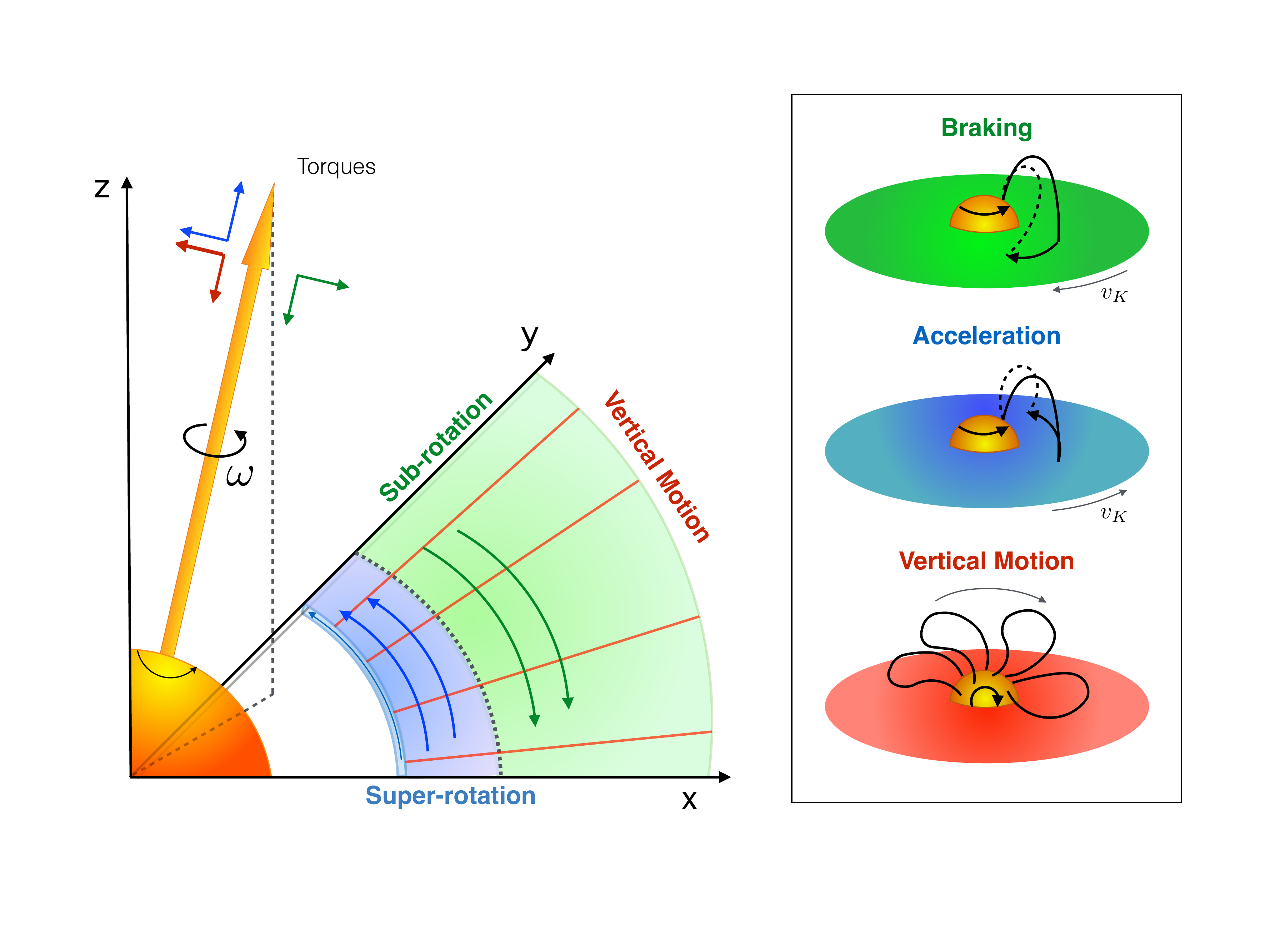}
\caption{A schematic to illustrate the origin of each magnetic torque. The blue region represents the disk material interior to corotation, including the inner wall of the disk, which super-rotates with respect to the stellar spin, acting both to spin the star up and realign its axis with the disk's. The green region is further out and so rotates more slowly, braking the stellar spin and acting to misalign the star. Red lines represent the requirement that the entire magnetosphere must be dragged vertically through the disk once per stellar rotation period if the star and disk are misaligned. A simple illustration of the physical mechanism behind each torque is shown on the right. The colored arrows in the top-left denote the net torque acting upon the stellar spin axis: green regions slow down and misalign the star, blue regions speed the star up and force realignment, whereas red regions act to brake stellar rotation whilst realigning the stellar spin-axis. Summed together, the resultant magnetic effect is usually to realign the disk and star with each other.}
\label{Torques}
\end{figure*}

The result is that as the star spins, the field loops are forced to bunch up on one face of the disk and become rarified on the opposed side of the disk (red diagram in Figure~\ref{Torques}), similarly to how water is pressurized on the leading edge of a boat paddle. Unlike the azimuthal induction, this paddle-like braking occurs over the entire disk as opposed to solely the region orbiting beyond co-rotation. As such, this effect constitutes a significant source of angular momentum loss for the star.

In addition to the intuitive braking torque upon stellar spin rate, the vertical component of magnetic induction leads to an additional torque affecting stellar orientation. In order to calculate the magnitude of these torques, we take a similar approach to that used within the azimuthal framework above. Specifically, we suppose that the component of the magnetosphere parallel to the disk is built up over half the disk and reduced on the other half to a degree which is of the same order as the stellar magnetosphere. Again, we suppose non-steady reconnection to provide a bound on magnetic field magnitude. Thus, we suppose the field induced by vertical field dragging (with subscript ``p") is given by

\begin{equation}
\mathbf{B}_{\textrm{p}}=\zeta(\phi)\big[B_r\,\mathbf{\hat{e}}_r+ B_\phi \mathbf{\hat{e}}_\phi\big],
\end{equation}

\noindent where $\zeta(\phi)=\sin(\phi-\psi)$. Recalling that $\psi$ is the azimuthal angle of the stellar spin pole direction projected onto the disk plane, this functional form for $\zeta$ ensures that field lines bunch up on the faces of the disk where magnetic field lines are being pushed into the disk and the field is rarified on the other faces, where the field is being rotated away from the disk surface\footnote{The $\zeta=\sin(\phi-\psi)$ form additionally ensures that $\nabla \cdot \mathbf{B_\textrm{p}}$=0.}.

 Finally, vertical field motion generates torques which take the form,
 
 \begin{align}\label{paddle}
 \tau_x &=-\frac{7\,\pi\,\mathcal{T}}{6\,a_{\textrm{in}}^3}\sin(\beta)^2\cos(\psi)\nonumber\\
 \tau_y &=-\frac{7\,\pi\,\mathcal{T}}{6\,a_{\textrm{in}}^3}\sin(\beta)^2\sin(\psi)\nonumber\\
 \tau_z &=-\frac{2\,\pi\,\mathcal{T}}{3\,a_{\textrm{in}}^3}\cos(\beta)\sin(\beta),\nonumber\\
 \end{align}
 
 \noindent where we have again invoked $a_{\textrm{out}}\gg a_{\textrm{in}}$. 
  
  We complete the specification of the torques arising from vertical field dragging by noting that if the star spins sufficiently slowly, the field lines will be able to diffuse vertically through the disk within one rotational period. Essentially, this is equivalent to saying that the field lines no longer reconnect and a true steady state is attainable (at least in terms of vertical field dragging). Suppose that the disk's diffusivity is prescribed as
  
  \begin{align}
  \eta=\bar{\alpha}_{\textrm{s}}h^2\,n_{\textrm{K}},
  \end{align}
  
  \noindent where $\bar{\alpha}_{\textrm{s}}$ is the Shakura-Sunyaev parameter for disk viscosity \citep{Shakura1973} and $n_{\textrm{K}}$ is the Keplerian angular velocity. We may suppose that the stellar field lines could diffuse vertically through the disk over a timescale $\tau\approx h^2/\eta=1/(\bar{\alpha}_{\textrm{s}}\,n_{\textrm{K}})$. Put another way, if the star is spinning more slowly than about $1/\bar{\alpha}_{\textrm{s}}$ times the angular velocity of the gas in the disk, we would expect field lines to diffuse through with greatly reduced magnetic induction. The likely value of $\bar{\alpha}_{\textrm{s}}$ within the inner disk ranges widely, from $10^{-1}$ to $10^{-3}$. Owing to this range, for definiteness we suppose that negligible field is induced if the star rotates less than about 10 times the Keplerian angular velocity of the disk. We include such an effect qualitatively by modulating the torques in equation~(\ref{paddle}) by a factor of $\exp[-(\omega/\epsilon\,\omega_0)^{-2}]$ with $\epsilon=0.1$ and $\omega_0$ corresponding to an $8\,$day spin-period.    
\subsubsection{Summary of Magnetic Torques} 

The cumulative impact of the magnetic torques derived above is depicted visually in Figure~\ref{Torques} and may be summarized as follows. The inner regions of the disk (i.e. $a'<a'_{\textrm{co}}$) rotate faster than the stellar magnetosphere, dragging field lines azimuthally, ahead of the stellar rotation. The result is both an acceleration of the stellar spin rate and a torque that acts to \textit{realign} the disk and stellar spin pole. Note that previous authors have concluded ambiguity over the sense of the tilting component to this torque \citep{Lai2011} largely owing to uncertain parameterizations for the stellar wind. Our analysis here demonstrates that the pure impact of the magnetic torques from the inner disk is realignment. As can be seen from equations~(\ref{Inner}), the $z$-component (i.e., disk-aligned) component is always positive, which tends to spin up and realign. The horizontal components, however are also positive and so tend to misalign, but simply adding the $z$- and horizontal components in quadrature results in a net realignment torque. 

The sense of torques is exactly reversed in the outer disk, i.e., the outer regions have a tendency to misalign. Torques arising from vertical induction tend predominantly to brake the stellar spin rate but they include a small component acting to realign. In order to complete the specification of magnetic torques, we must choose a value for the inner disk radius $a'_{\textrm{in}}$. In our models, we adopt a so-called ``disk-locked" configuration \citep{Konigl1991,Mohanty2008}, whereby the outer and inner disk torques cancel at some prescribed stellar period (which we set to 8\,days; see section~(\ref{Spin}) below). Accordingly, the inner disk truncation radius lies at $\sim0.8\,a_{\rm{co}}|_{\beta=0}$, in agreement with more sophisticated numerical simulations \citep{Long2005}.

\subsection{Stellar Spin Rate}\label{Spin}

A complete account of the factors influencing the spin rates of young stars remains elusive \citep{Matt2004,Herbst2007,Littlefair2010}. The reason is that if one na\"ively assumes a T-Tauri star to contract along its Hyashi track whilst conserving angular momentum, it would be expected to spin up sufficiently quickly such as to reach break-up velocity within the disk lifetime. However, observations clearly demonstrate that the vast majority of T-Tauri stars spin at rates far reduced from their break-up angular velocity. Specifically, most young stellar objects rotate at about $3-10$\,day periods with only a rare few which are indeed close to break-up angular velocity, at periods of $\sim1$\,day. The slow spins of such stars is sometimes referred to as the `stellar angular momentum problem' because stars spin slower than they ``should". 

It has long been suspected that magnetic star-disk interactions play a key role in modulating the spin rates of T-Tauri stars (see \citet{Bouvier2014} for a recent review). Here, we do not attempt any sophisticated modeling aimed at explaining the stellar spin rates in detail. However, we note that the quasi-periodic twisting and opening of magnetic field lines discussed above provides a conduit through which angular momentum may be lost from the system, by way of stellar and/or disk winds. In addition, the torque arising from vertical field motion, derived above, provides an extremely efficient braking mechanism upon stellar spin, provided the stellar magnetosphere is sufficiently strong. As such, mutual inclination between a star and its disk may give rise to hitherto under appreciated magnetohydrodynamic influences upon stellar spin, especially when considering stars with lower masses ($M\lesssim1.2\,\textrm{M}_{\odot}$) and strong magnetospheres ($\sim$1\,kGauss).

Within the framework of our model, we treat spin-rate evolution as follows. We allow the torques prescribed above to act freely on stellar spin rate. In addition, we introduce a relaxation factor, reflecting the reluctance of stars to spin up to break-up, whereby the star seeks the disk-locked spin rate $\omega_{\textrm{r}}$ over a timescale $\tau_{\textrm{r}}$. The relaxation is prescribed as,

\begin{align}
\bigg(\frac{d\omega}{dt}\bigg)_{\textrm{r}}=-\frac{\omega-\omega_0}{\tau_{\textrm{r}}},
\end{align}

\noindent where we choose, for $\tau_{\textrm{r}}$, the stellar contraction timescale, which may be derived from equation~(\ref{Contract}):

\begin{align}
 \tau_{\textrm{r}}\equiv (\dot{R}/R)^{-1}=\frac{24\pi}{5-\xi}\frac{R_{\star}^3\sigma T_{\textrm{eff}}^4}{G M_{\star}^2}.
\end{align}

\noindent  The above timescale changes as the star contracts, but remains of order $\sim1$\,Myr which is the value we adopt in this work. As already stated above, we choose the equilibrium angular velocity to be within the observed range, $\omega_{\textrm{0}}=2\pi/(8\,\textrm{days})$.

\subsection{Frame of Reference}

It is beneficial to carry out all calculations in the frame of a distant, binary companion to the central star. As such, we follow the approach of \citet{Peale2014} and define Euler angles within the binary frame related to the nutation, precession and rotation of the rigid body while assuming exclusively principal axis rotation (this is an excellent approximation for a T-Tauri star, spinning at a period of $3-10\,$days). Specifically, $\tilde{\beta}$ is the angle between the central star's spin axis and the binary orbit normal; $\tilde{\Omega}$ is the longitude of ascending node of the star in the binary frame where $\tilde{\Omega} = 0$ implies commensurate disk and stellar lines of nodes; and the third Euler angle $\varphi$ is the angle through which the star rotates as it spins ($\varphi$ only enters the equations as a rate of change: $\dot{\varphi} = \omega$). 

The equations for the evolution of $\tilde{\beta}$, $\omega$ and $\tilde{\Omega}$, adapted from \citet{Peale2014} are:

\begin{align}
\label{pealeequns}
\frac{d\omega}{dt}&=-\frac{\omega-\omega_0}{\tau_{\textrm{r}}}+\sin(\tilde{\beta})\bigg[N_{\bar{x}}\sin(\tilde{\Omega})\nonumber\\
&-N_{\bar{y}}\cos(\tilde{\Omega})\bigg]+\cos(\tilde{\beta})\,N_{\bar{z}},\nonumber\\
\frac{d\tilde{\beta}}{dt}&=-\frac{1}{\omega} \bigg[\cos(\tilde{\beta})(-N_{\bar{x}} \sin(\tilde{\Omega})+N_{\bar{y}} \cos(\tilde{\Omega}))\nonumber\\
&+N_{\bar{z}} \sin(\tilde{\beta}))\bigg],\nonumber\\
\frac{d\tilde{\Omega}}{dt}&=-\frac{1}{\omega\, \sin(\tilde{\beta})}\bigg[N_{\bar{x}} \cos(\tilde{\Omega})+N_{\bar{y}} \sin(\tilde{\Omega})\bigg],
\end{align}

\noindent where $N_{\bar{i}}$ are projected torques. Note that by fixing the disk's longitude of ascending node at $\Omega' = 0$, we have implicitly placed ourselves into a frame co-precessing with the disk's angular momentum vector, as discussed above. The effect of precession was included within the gravitational part of the calculation (equation~(\ref{HammyK})) and so we need not retain it here.

The projected quantities $N_{\bar{i}}$ are directly related to the torques calculated above, although the components of the torques in the disk frame, $-\tau_{i'}$, must first be projected onto the Cartesian axes in the binary frame. Such a projection constitutes a simple geometric rotation of co-ordinates because the disk-binary inclination is a constant of motion. The co-ordinate rotation transformation angle is fixed at some prescribed value, $\beta'$ to the $z$-axis and its sense is defined as anti-clockwise about the $x$-axis. As such, the components, $N_{\bar{i}}$ are given in terms of $\tau_i$ by:

\begin{align}
N_{\bar{x}}&=-\tau_{x'}/(I M_{\star} R_{\star}^2),\nonumber\\
N_{\bar{y}}&=-(\cos(\beta')\tau_{y'}-\tau_{z'} \sin(\beta'))/(I M_{\star} R_{\star}^2),\nonumber\\
N_{\bar{z}}&=-(\cos(\beta')\tau_{x'}+\tau_{y'}\sin(\beta'))/(I M_{\star} R_{\star}^2).
\end{align} 

\noindent The above equations can be used to model the dynamics of the central star resulting from its magnetic interactions with its protoplanetary disk whilst the disk itself precesses within the binary frame.

\section{Results}

\begin{figure}[h!t]
\centering
\includegraphics[trim=6cm 4.8cm 4cm 6cm, clip=true,width=1.1\columnwidth]{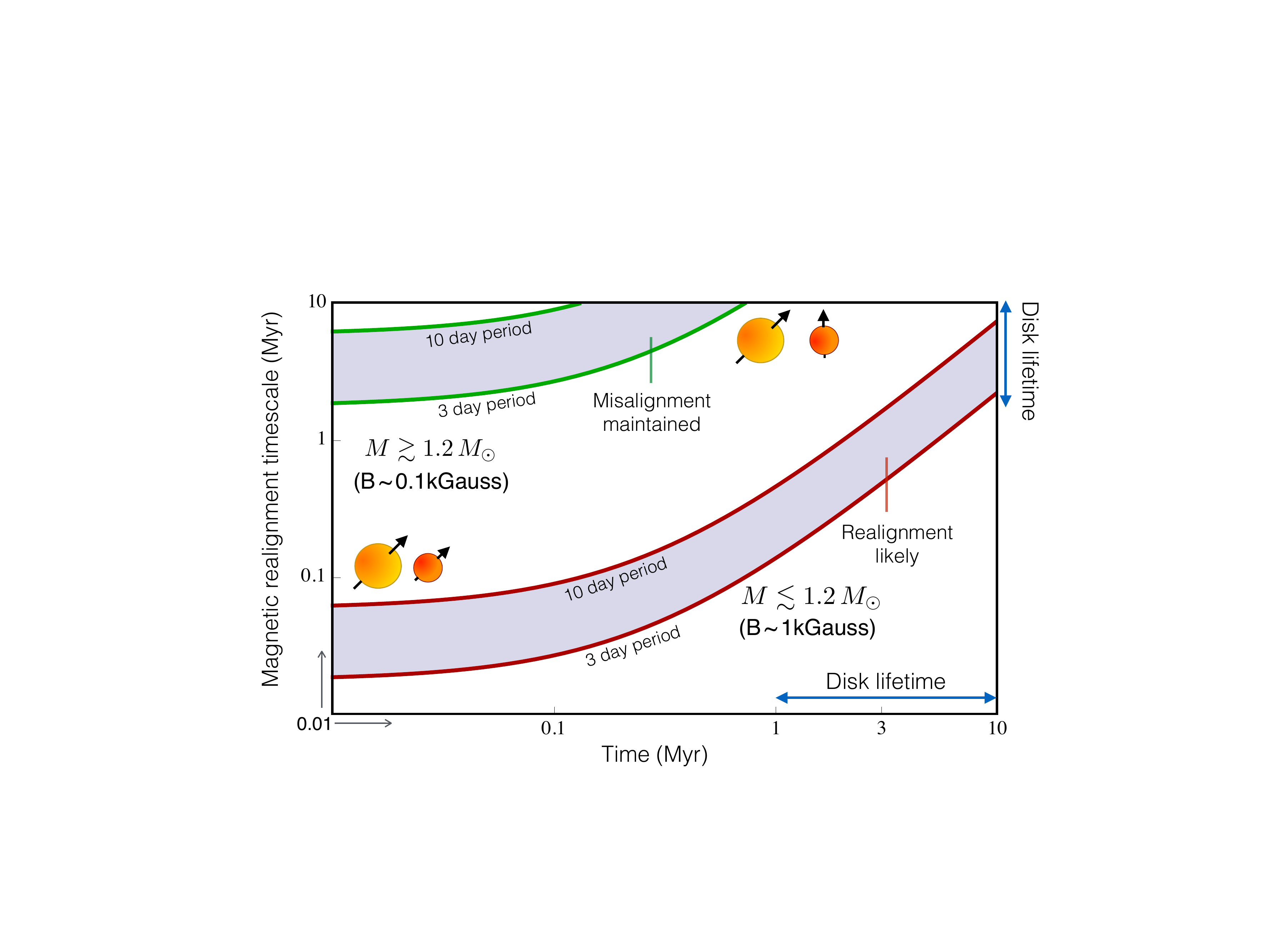}
\caption{The approximate magnetic torquing timescale ($T_{\textrm{align}}$) as a function of disk-star age for four regimes. The green lines apply to high-mass stars with a surface field strength of $\sim0.1$\,kGauss. Red denotes low-mass stars with a surface field strength of $\sim1$\,kGuass. In both cases, the upper line considers the timescale relevant to a star which is spinning with a 10 day period whereas the lower line applies to one with a 3 day period. The stellar spin rate is assumed to result from the locking to a circumstellar disk \citep{Konigl1991} and so the faster-spinning cases consider a disk which is truncated at smaller radii, increasing the magnetic influence. Notice that the dominant effect upon magnetic torquing timescale is the magnetic field strength, with timescales proportional to its inverse square. The timescales increase with time because the star contracts, leading to an effectively weaker field at the position of the inner disk. Only for the very earliest stages of protoplanetary disk evolution are high-mass stars' magnetospheres strong enough to significantly alter their orientation whereas low-mass stars remain dynamically influenced by magnetic fields throughout the entire typical disk lifetime of $\sim1-10$\,Myr.}
\label{Timescales}
\end{figure}

We motivate the following analysis by considering the timescales over which the magnetic torques act, as derived above. Specifically, magnetic torques increase as $\propto B^2$, meaning that the order-of-magnitude difference in field strengths between high and low-mass stars translates to an enhancement of two orders of magnitude in the torques felt by lower-mass stars. Supposing the star to be set up in a retrograde, anti-aligned state ($\beta=\pi$), we calculate the characteristic realignment timescale $T_{\textrm{align}}$ using the equations derived above:
\begin{align}
T_{\textrm{align}}\equiv \frac{\omega}{\dot{\omega}}=\frac{3 G M_{\star}^2 I}{4 \pi \omega}\frac{\mu_0}{B_{\star}^2R_{\star}^4}
\end{align}
where $R_{\star}$ follows the time-dependence described in equation~\ref{Rstar}, giving rise to a time dependent magnetic torquing timescale which we illustrate in Figure~\ref{Timescales}. Taking $1$\,kGauss as the field strength typical of low-mass stars, under nominal star-disk parameters their absolute re-alignment timescales are of the order of $\sim1$\,Myr throughout the majority of the disk lifetime. Conversely, the analogous timescale for the 0.1\,kGauss fields typical of high-mass stars is closer to $\sim100$\,Myr (Figure~\ref{Timescales}). These timescales are respectively shorter and longer than the typically-quoted $3-10\,$Myr lifetimes of protoplanetary disks \citep{Haisch2001}. Therefore, magnetic interactions only have the potential to wipe out primordial star-disk misalignments in low-mass systems.

\subsection{A companion inclined by 30\,degrees}

\begin{figure*}[h!t]
\centering
\includegraphics[trim=4.8cm 5cm 2cm 4.3cm, clip=true,width=1\textwidth]{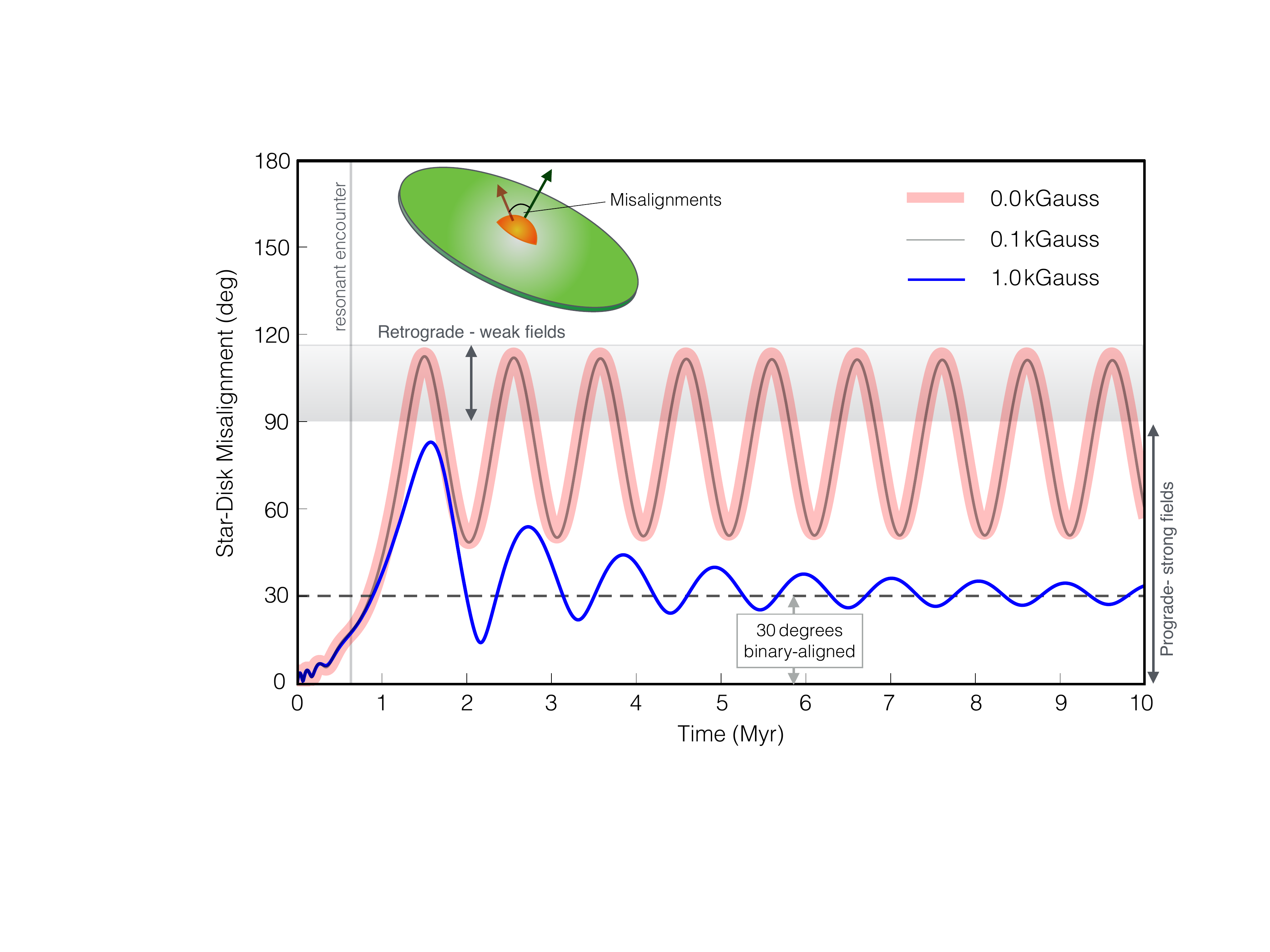}
\caption{The time evolution of mutual star-disk inclination. We consider a binary companion to orbit the system at an inclination of 30\,degrees relative the disk (greater angles are displayed in Figure~\ref{MainGraph2}). The companion is prescribed to cause the disk to precess with a 1\,Myr period. The purely gravitational case is shown as a thick, pink line. The thin, black line denotes evolution in the presence of the weak fields of high-mass stars ($\sim$\,0.1\,kGauss) and the blue line denotes the evolution corresponding to the strong fields of low-mass stars ($\sim$\,1\,kGauss). As in previous work \citep{Spalding2014}, we find that a secular spin-orbit resonance is encountered as the disk loses mass and the star contracts. However, significant misalignments are inhibited for the stronger magnetic fields characteristic of low-mass stars whereas the fields of high-mass stars make no appreciable difference to the dynamics. Interestingly, the action of magnetic torques takes the star into alignment with the binary plane, not the disk plane, suggesting that small misalignments are indeed a natural outcome of low-mass star evolution, whereas high-mass stars can take on the full range of misalignments, in direct agreement with the observations.}
\label{MainGraph}
\end{figure*}

In order to place magnetic realignment into the disk-torquing context, we integrate the mutual star-disk inclination over a 10\,Myr timescale (i.e. an upper bound on typical disk lifetimes; \citealt{Haisch2001}), adopting the case of a 1\,M$_{\odot}$ massive companion inclined at 30\,degrees to the disk plane. We present three cases in Figure~\ref{MainGraph}: the purely-gravitational scenario (thick, pink line), the dynamics of a star with a dipole magnetic field strength of 0.1\,kG, characteristic of higher-mass stars (thin grey line) and the situation for a dipole field strength of 1\,kG, typical of low-mass stars (blue line). 

Both the magnetic field-free and 0.1\,kG cases look almost identical. In other words, the magnetic fields of higher-mass stars are dynamically unimportant within the disk lifetime, in accordance with the timescale analysis quoted above. The gravitational evolution is then equivalent to that discussed in \citet{Spalding2014}. Specifically, at the earliest times, the star precesses about the disk plane much faster than the disk precesses about the binary plane (the high-frequency wiggles at time $t\lesssim0.3$\,Myr) and small misalignments are maintained. However, as the star shrinks and the disk loses mass, gravitational star-disk coupling weakens until the two precession timescales are roughly commensurate. This situation causes the system to pass through a secular resonance \citep{Murray2000,Morby2002}, facilitating a brief period of extremely efficient angular momentum transfer between the disk and star, resulting in large misalignments. 

Importantly, even with the initial companion-disk inclination set to 30\,degrees, a retrograde disk may emerge from the secular resonance. It is this non-linearity between disk-binary inclination and resulting misalignment which is the essence of how disk-torquing can account for the entire range of observed misalignments \citep{Spalding2014}.

The same secular resonance is encountered in the 1\,kG case, but crucially, the resulting large misalignments are erased within typical disk lifetimes. An important aspect of the dynamics is that the orientation of the star does not converge upon the disk plane, but rather, it converges to 30\,degrees, i.e., the binary plane., which may be explained qualitatively as follows. Gravitational systems are conservative, allowing their dynamics to be described as following contours of a scalar function, or Hamiltonian \citep{Morby2002}. Magnetic torques introduce a dissipative component that acts to turn the elliptical equilibrium points of the Hamiltonian into attractors. In this case, the attractor is the binary-aligned state.

A crucial point to emphasize is that \textit{in aligning with the binary plane, retrograde disks are prohibited}. To see this, note that the companion's gravitational perturbation upon the disk is dynamically equivalent whether the binary orbit is clockwise or counter-clockwise (as defined in some arbitrary frame). Such symmetry arises from approximating the orbit as a massive wire and has the consequence that the equilibrium of the Hamiltonian never lies at a position where the star is tilted by more than 90$\,\deg$ to the disk.

\subsubsection{Sensitivity to Disk-Binary Inclination}

\begin{figure*}[h!t]
\centering
\includegraphics[trim=0cm 2cm 0cm 0cm, clip=true,width=1\textwidth]{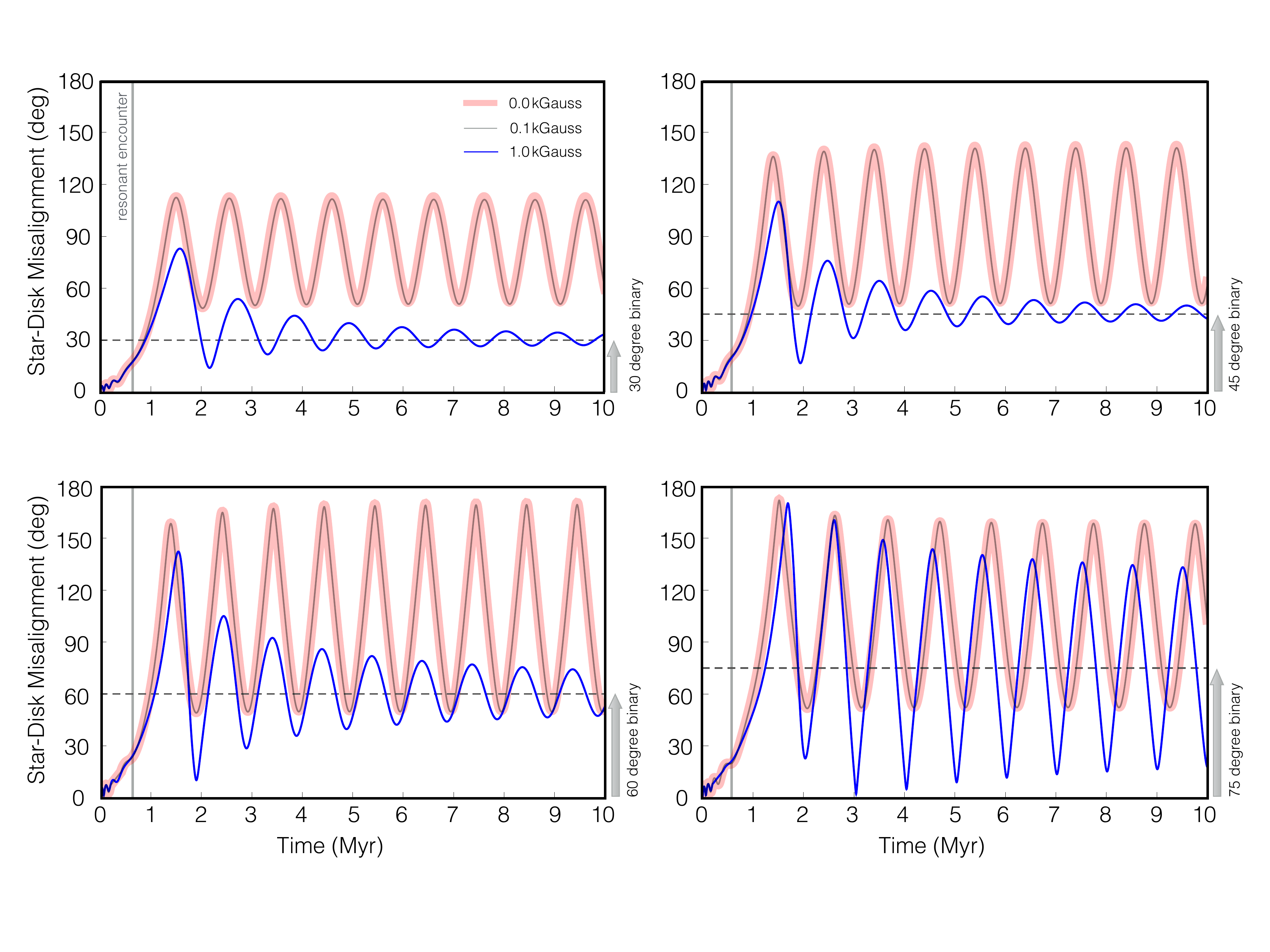}
\caption{Star-disk misalignments as functions of time for a variety of disk-binary inclinations. This set of figures show similar information to that shown in Figure~\ref{MainGraph}, except we now illustrate the evolution for a range of angles: $30\deg$ (top-left), $45\deg$ (top-right), $60\deg$ (bottom-left) and $75\deg$ (bottom-right). The thick, pink line is the field-free case, the thin, black line presents the weak-field (0.1\,kGauss) case inherent to high-mass stars and the blue line denotes the strong-field (1\,kGuass) case of low-mass stars. The over-all pattern is largely similar up to $60\deg$, in that the star is drawn towards a binary-aligned state over the magnetic realignment timescale. Not even the strong fields can undo the extreme resonant acquisition of misalignments occurring as a result of a $75\deg$ binary inclination. These larger binary inclinations are less likely, but raise the possibility that we may find a rare population of retrograde planetary orbits around low-mass stars in future datasets. }
\label{MainGraph2}
\end{figure*}

For completeness, we investigate the star-disk dynamics over a range of disk-binary inclinations in addition to the 30\,deg case above. Specifically, Figure~\ref{MainGraph2} illustrates the more extreme cases of 45, 60 and 75\,degree inclinations. The picture is very similar across 30, 45 and 60\,degrees, i.e., the magnetic fields of low-mass stars are capable of realigning them with the binary plane within the disk lifetime for a broad range of angles. In contrast, even the stronger fields were unable to wipe out retrograde disks when an extreme initial binary-disk inclination of 75\,degrees is chosen. As expected, the orientation of the star is almost entirely unaltered by magnetic disk-star interactions for $0.1$\,kG fields in all cases. 

\begin{figure}[h!t]
\centering
\includegraphics[trim=9cm 7cm 6cm 6cm, clip=true,width=1\columnwidth]{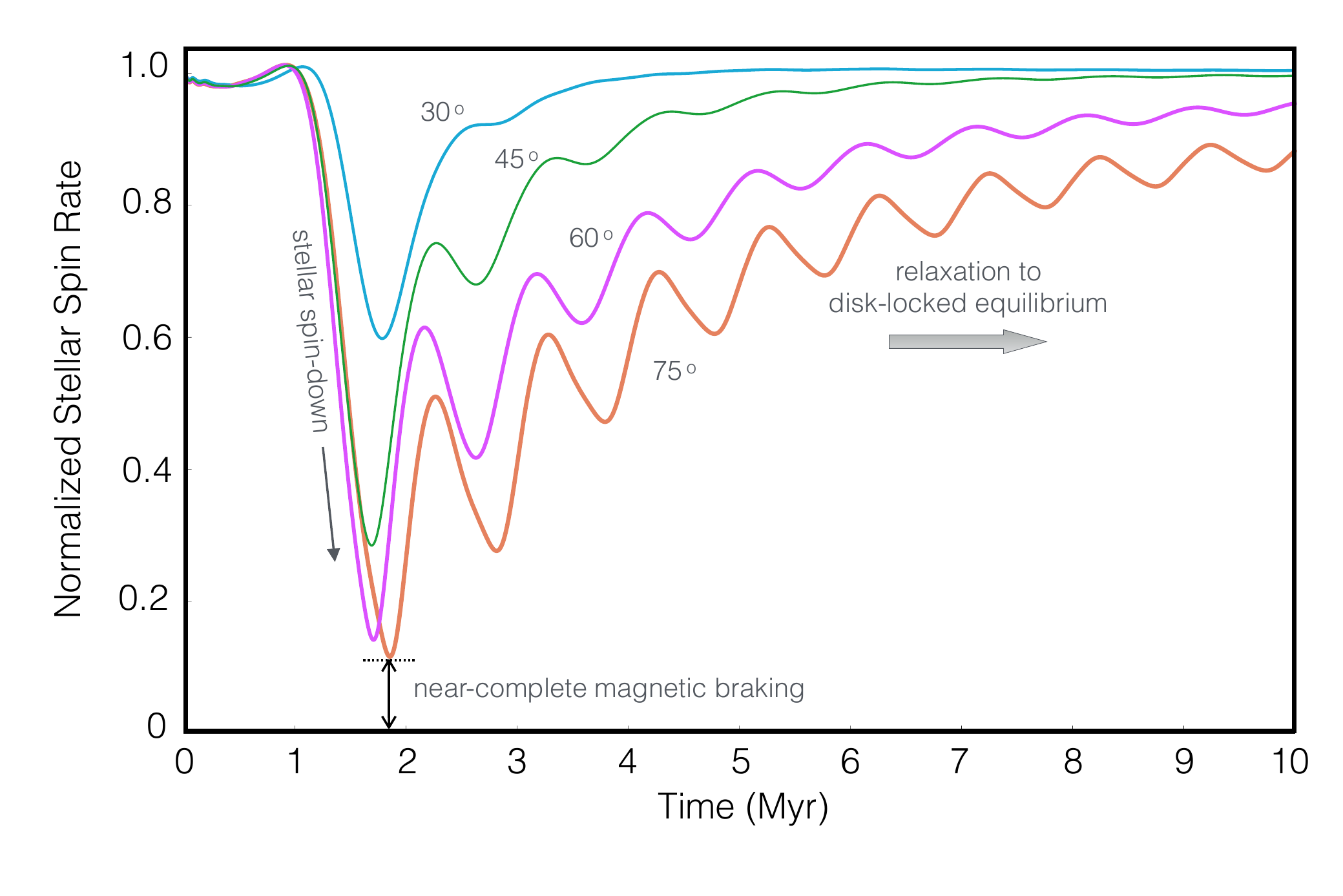}

\caption{The time evolution of absolute stellar angular velocity in the case of a 1\,kGauss dipole field, plotted in units of the equilibrium angular velocity (here corresponding to an 8\,day period). We show the evolution for four values of disk-binary inclination, increasing from top to bottom: $30\deg$ (cyan line), $45\deg$ (green line), $60\deg$ (purple line), $75\deg$ (orange line). In each case, the star is prescribed to relax to a disk-locked equilibrium at an 8\,day rotational period over a Kelvin-Helmholtz time. The relaxation is included in an \textit{ad hoc} fashion and so the exact form of the rotation curves after $\sim2\,$Myr is not to be taken too literally. However, as is most apparent for the higher inclinations, magnetic braking constitutes a significant mechanism for the removal of stellar angular momentum. Indeed, for large binary inclinations, the star can be almost entirely stopped and re-spun within a relatively brief time interval.}
\label{MainGraph3}
\end{figure}

We also present the time evolution of stellar spin resulting from the dynamics. A notable effect of larger binary inclinations is the considerable impact upon stellar spin rate, displayed in Figure~(\ref{MainGraph3}). Though one should not take the rotation periods displayed in Figure~(\ref{MainGraph3}) too literally, owing to the uncertainties in prescribing rotation rates, an important aspect of the set-up emerges. Specifically, tilted stars will interact with their disks in such a way as to significantly brake stellar rotation, with such braking being more significant for larger star-disk inclinations. The origin of such an effect is in the requirement of a tilted star to drag its entire magnetosphere vertically through the disk once every period\footnote{Note that ignoring the torques arising from vertical field motion would actually predict spin-\textit{up} of a tilted star owing to the expanded corotation radius.}. For high binary inclinations, the star is potentially spun down to nearly a stand-still as the system passes through the secular resonance, before being spun-up again in the direction of the disk.

\section{Discussion}

\subsection{Summary}

Prior to this work, a deficiency of the ``disk-torquing" model \citep{Batygin2012,Batygin2013,Spalding2014} for the acquisition of spin-orbit misalignments lay in its inability to reproduce the observed mass-dependence of misalignments \citep{Winn2010,Albrecht2012}. Here, we resolved this discrepancy through the addition of a comprehensive set of magnetic star-disk torques on top of the previously-considered gravitational dynamics. Taking account of the recently-observed mass-dependence of T-Tauri dipole field strengths \citep{Gregory2012}, the observed trend arises naturally. 

We began by deriving analytical expressions describing the magnetic star-disk torques. Many previous authors have contributed to the development of a description of such torques \citep{Ghosh1978,Livio1992,Armitage1996,Agapitou2000,Uzdensky2002,Matt2004,Lai2011}. A common limitation has been that torques are considered to arise solely as a result of magnetic induction associated with relative \textit{azimuthal} motion between the stellar magnetosphere and the disk fluid. However, here also considered magnetic induction arising from oblique rotation. To this end, we have found that the associated torques provide an important correction to the existing pictures.  Accordingly, this effect constitutes an efficient source of braking upon the stellar spin.  

 The largest contribution to the torques affecting stellar orientation, as opposed to spin rate, arise from azimuthal induction. The influence of the disk material inside corotation is to spin the star up and align the stellar spin pole with the disk, with the opposite effect arising from the outer disk. This effect is amplified by stellar obliquity. In other words, when the star is tilted, a greater proportion of the disk is acting to accelerate the stellar spin and align it with the disk. Inclusion of this aspect increases the stability of the disk-locked equilibrium proposed elsewhere \citep{Konigl1991,Mohanty2008}.
 
 The final addition to our picture of magnetic torques was to include (to leading order) the effect of finite disk thickness. Whilst simple and widely adopted, a short-coming of the razor-thin disk model is that a star inclined by 90\,degrees to the disk is predicted to feel no magnetic torques arising from azimuthal field dragging alone (vertical induction still occurs). This is an artifact that comes about by supposing that there exists a negligible solid angle within which field lines penetrate radially the disk. However, though disk aspect ratios are small ($\sim 0.1$), the inner edge of the disk is the closest point on the disk to the star and therefore can communicate a non-negligible torque to the central star (Equation~\ref{Inner}).

We motivated the potential for magnetic torques to sculpt the mass-dependence of obliquities by showing that the calculated magnetic torquing timescale is $\sim$1\,Myr for the 1\,kG fields typical of low-mass stars whereas the analogous timescale applicable to high-mass stars was closer to 100\,Myr. The fact that a typical protoplanetary disk lifetime sits right in the middle of these two timescales, at 3-10\,Myr, indicates qualitatively different evolutionary scenarios during the disk-hosting stages of high and low-mass stars. In other words, star-disk magnetic torques constitute an influential factor in low-mass systems but may be neglected in high-mass systems.

The timescale analysis is compelling, and demonstrates that, were one to place a star and its disk in a misaligned configuration and leave them alone, a 1\,kG dipole field would realign them before the disk dissipates. However, stars are likely to be aligned with their disks in the absence of outside influences \citep{Spalding2014b}. The excitation of misalignments must then come about by way of perturbations arising from a companion star \citep{Batygin2012}. Appropriately, next we coupled the updated magnetic torques derived here to the purely gravitational model presented in \citet{Spalding2014}. Specifically, the star and the disk interacted through both gravitational and magnetic torques, with a massive companion causing the disk to precess with a period of 1\,Myr.

As expected from the timescale analysis, the 0.1\,kG fields characteristic of high-mass stars resulted in little deviation from a purely gravitational picture. However, the stronger 1\,kG fields caused the star to align, not with the disk, but with the binary plane (except for extreme disk-binary inclinations of $>$75\,degrees). Thus, the most important result of our paper emerges: \textit{Dipole fields typical of lower-mass T Tauri stars realign the stellar spin axis with the plane of a perturbing companion on a timescale shorter than the disk lifetime. In this way, small misalignments occur naturally, but retrograde disks are prohibited in all but the more massive systems.} 

Our results agree with the observational data presented in Figure~\ref{Lambda}. Retrograde circular orbits only exist among the higher-mass ($M\gtrsim$\,1.2\,M$_\odot$) population of stars. Despite the absence of retrograde orbits around the lower-mass stars, significant obliquities still persist in general. It is important to note, also, that the observations only reveal the \textit{projected} obliquity of exoplanetary systems. A true inclination of 60\,degrees, say, will on average appear sky-projected as a lower obliquity within our current observational dataset. On the other hand, distinguishing retrograde from prograde orbits is a much simpler task. The lack of retrograde systems, together with the degeneracy inherent to measuring sky-projections, means that the current observations are fully consistent with our analysis here whereby small alignments persist while retrograde orbits are prohibited.

A more subtle point is that, although the low-mass inclinations approach binary-aligned equilibria, their time-evolution traces an oscillatory trajectory. Such oscillations in stellar orientation, especially for the 60 and 75\,degree cases (Figure~\ref{MainGraph2}), take the star retrograde with respect to the disk for a small portion of the trajectory. If the disk were to rapidly dissipate before the amplitude of oscillations was sufficiently damped by magnetic interactions, and at a point in the oscillations where the star was in a retrograde position, such a system could conceivably produce retrograde planetary systems around low-mass stars. This scenario is much less likely than the alternative result of prograde systems, but does allow for the future detection of retrograde systems around low-mass stars, albeit at a significantly reduced frequency as compared to prograde systems.

Our analysis was largely based upon order-of-magnitude estimates. In accordance with this level of precision, we had to make several assumptions about the star-disk configuration. Perhaps most importantly, we chose the inner truncation radius of the disk such as to mimic a ``disk-locked" scenario whereby the inner and outer disk torques cancelled at a specified equilibrium, aligned spin rate. In accordance with observations \citep{Bouvier2013}, we decided upon 8\,days for this equilibrium spin period in our simulations. As can be seen in Figure~(\ref{Timescales}), the chosen equilibrium period does indeed influence the characteristic magnetic torquing timescale. Had we chosen 3\,days, for example, disk-locking would require a smaller truncation radius, resulting in stronger disk-star torques (both magnetic and gravitational). However, the relative variation in timescales between 3 and 10\,day configurations is far smaller than the difference between high and low-mass stars (owing to the $B^2$ dependence). Furthermore, because 8\,days is in the middle of these extremes, the end result with respect to star-disk inclination is qualitatively similar across all realistic star-disk configurations.

One final omission that is worth a brief mention is the potential argument that, because our picture (disk-torquing) requires the presence of a companion, perhaps the mass-misalignment trend is merely a reflection of higher binarity in higher-mass stars. Whilst this particular aspect has not been examined in depth, \citet{Crida2014} analyzed the expected spin-orbit distribution arising out of the purely-gravitational disk-torquing picture presented in \citet{Batygin2012}. Whilst the expected distribution was consistent with that observed it was difficult to simultaneously fit the highest and lowest obliquities and little attention was paid to any mass dependence. It is difficult to see how such a sharp transition in obliquities may arise purely from such a smooth occurrence relation as mass vs binarity and so we feel the much sharper nature of the dipole field strength transition is a more likely cause.

\subsection{Viability of the Disk-Torquing framework}

Ultimately, our work here is a demonstration that the observed spin-orbit misalignments in exoplanetary systems can come about predominantly by way of mechanisms occurring during the early, disk-hosting phase. We do not expect the entirety of the observed misalignments to have originated through disk-torquing because dynamical interactions probably play a secondary role in the hot Jupiter delivery process \citep{Ford2006}. It is important, therefore, to contrast the predictions of each model in order to determine which mechanism, if any, is dominant. 

Dynamically-excited inclinations are likely to coincide with high eccentricities. Although controversy exists between analytic and numerical analyzes over the sense of eccentricity evolution during disk-driven migration (e.g. \citealt{Goldreich2003,Bitch2010,Kley2012}), the general consensus is that eccentricity is limited to small values. As such, we predict that as datasets become more complete, the mass-misalignment trend will remain for circular systems, but not for eccentric systems ($e\gtrsim0.1$). Such a pattern is already beginning to emerge (Figure~\ref{Lambda}) but will be fully tested when the upcoming \textit{TESS} mission commences. Furthermore, a disk-torquing origin for misalignments is fully consistent with the existence of multi-transiting systems \citep{Huber2013}. Indeed, we expect future observations of such systems to reveal the same mass-misalignment trend as that currently measured for single planet systems, with the break between high and low obliquities occurring at a similar (although not necessarily identical) stellar mass.

Our work here has focussed largely on close in systems. What can be said about more distant planets? An implicit assumption we have made is that throughout the dynamics, the disk acts like a rigid, planar body. Numerical simulations have demonstrated that in reality, the disk is likely to develop a mild warp in response to the perturbations of a companion \citep{Larwood1996}. Therefore, the possibility exists for the outer planets to occupy a different plane from the inner planets despite having undergone disk-torquing.  

A more subtle difference between the outer and inner regions exists. One might imagine that close-in planets, having migrated through their natal disks, will be much more shielded from dynamical instabilities arising from perturbations later in the systems lifetime. This shielding arises both as a result of sitting in a deeper gravitational potential well, but also through the stabilizing influence of general relativity\footnote{Indeed, general relativity provides a considerable boost to stability of Mercury's orbit, despite being situated 10 times further from the Sun than some hot Jupiters are from their stars \citep{Batygin2015}.}. Accordingly, we might expect there to exist a population of close-in systems which have obtained their orbits by way of coplanar disk migration, in addition to a separate population of planets around the same star that, for whatever reason, remained at larger radii within their disk, becoming subject to various dynamical instabilities later in their evolution.

In conclusion, the magnetically-facilitated realignment presented here provides a natural pathway for the generation, not only of high obliquities, but also their observed dependence upon mass. Such a ``disk-torquing" model appeals to no assumptions beyond those within the bounds of what has been observed for T-Tauri systems, namely magnetic field strength, stellar spin rate, disk ionization state and disk mass. Out of such nominal parameters, the observed mass-misalignment trend arises naturally. Additionally, the conclusions presented here are fully consistent with the decades-old picture of smooth migration through a protoplanetary disk \citep{Goldreich1980}, leading to short-period orbits coplanar with the disk. 

\textbf{Acknowledgements} We would like to thank Fred Adams, Heather Knutson and Dave Stevenson for enlightening conversations. CS acknowledges funding from the NESSF15 Graduate Fellowship in Earth and Planetary Science. We would additionally like to thank an anonymous reviewer, whose comments led to an improved manuscript. This research is based in part upon work supported by NSF grant AST 1517936.

 \end{document}